\documentclass[a4paper,12pt]{article}
\usepackage[utf8]{inputenc}
\usepackage{cancel}
\usepackage{ulem}
\usepackage{amsfonts}
\usepackage{amssymb}
\usepackage{graphicx}
\usepackage{amsmath}
\usepackage{enumerate}
\usepackage{mathtools}
\usepackage{subfig}
\usepackage{color}
\usepackage{tikz}
\usepackage{float}
\usepackage{here}
\usepackage{cite}
\usepackage{mathrsfs}
\usepackage{float,epsfig}
\usepackage{dcolumn}
\usepackage{lscape}
\usepackage{subfig}
\usepackage{graphicx}
\usepackage{bm}
\usepackage{amsmath,amssymb,amsthm}
\usepackage[colorlinks=true,linkcolor=blue,citecolor=red]{hyperref}
\usepackage{multirow}

\usetikzlibrary{calc}
\newcommand{\be}{\begin{equation}}
\newcommand{\ee}{\end{equation}}
\newcommand{\bea}{\setlength\arraycolsep{2pt} \begin{eqnarray}}
\newcommand{\eea}{\end{eqnarray}}

\def\0{{\sst{(0)}}}
\def\1{{\sst{(1)}}}
\def\2{{\sst{(2)}}}
\def\3{{\sst{(3)}}}
\def\4{{\sst{(4)}}}
\def\5{{\sst{(5)}}}
\def\6{{\sst{(6)}}}
\def\7{{\sst{(7)}}}
\def\8{{\sst{(8)}}}
\def\sst#1{{\scriptscriptstyle #1}}

\makeatletter \@addtoreset{equation}{section}

\begin{document}

\title{{\normalsize \textbf{\Large  On Thermodynamics  of   Charged  Black Holes via  Extended  Space-time Derivatives  }}}
\author{  \small Adil  Belhaj, Maryem  Jemri \thanks{
Authors are listed  in alphabetical order.} \footnote{maryem.jemri@um5r.ac.ma} \hspace*{-8pt} \\
{\small ESMaR, Faculty of Science, Mohammed V University in Rabat, Rabat, Morocco  } }
\maketitle

\begin{abstract}
 Inspired by non-commutative geometry in  string theory, we propose extended derivatives in black hole physics   by incorporating  a real antisymmetric tensor of rank  2 carrying   similarities  of certain  stringy fields.  Using   gauge theory formulation of gravity via de Sitter group theory, we  first  find  the associated black hole solutions by solving  the  Einstein  field equations. Then, we  study  the thermodynamic  properties by approaching the stability analysis,  the criticality,  and  the phase transitions. Concretely, we investigate  the  $P$-$V$ criticality behavior of the obtained solution. We compute and   examine  the Gibbs free energy  revealing  
 comparable attitudes   with the Van der Waals phase transitions.  Combining such results, we provide constraints on the deformed parameter  $B$  and the charge $Q$ with the  help of CUDA numerical methods  exploited  in machine learning computations. Precisely, we show that  there are suitable ranges for  such  parameters where the obtained black holes behave like  the Van der Waals fluid   systems.
{\noindent} 

\end{abstract}

%

\newpage


\newpage

\section{Introduction}

Black holes in non-trivial physical theories are fascinating subjects that have received considerable 
attention. Such  objects, being governed by Einstein's general theory of relativity,  are linked to a region of the space-time being so massive with a gravitational field that even light cannot escape it due to the enormous gravity it exerts  \cite{2,3,4}.  The study of these objects has led to a better understanding of the fundamental laws of physics, including  quantum mechanics \cite{5}. They also shed light on the nature of the universe and the mysterious phenomena that occur within it.  It turns out that the phenomenon of  the black holes is of great importance, not only in astrophysics, but also in the
theories of high-energy physics, including string theory and related topics  that have been explored to combine  general relativity and quantum physics  \cite{6}. The study of black holes has been developed via two main  topics: thermodynamics and optics \cite{23,24,25,26,27,270,271,28,29,30,31,32,33}. Concerning thermal properties,  the  stability,   the criticality, and   the phase transition behaviors  have been examined 
 and inspected.   In optical terms,  the shadow and  the light deflection have been studied,  leading to interesting results that have been published in several papers.
 
 Recently, a special emphasis has been put on the investigation of   the   black hole properties from  non-trivial  space-time geometries by incorporating extra  deformation parameters.   In this way,   certain black holes  have been constructed via a gauge theory formulation of gravity  where the    de Sitter group  in four dimensions has been considered as   a local    gauge symmetry \cite{8}.  Combining such a gravity gauging model  with  space-time deformation  scenarios,  certain black holes  have been built. Concretely, the  black holes on   a non-commutative space-time  have been investigated by computing  relevant geometric quantities being  primordial in the   gravitational  computations relaying on the Einstein  field equations.  More recently,  a deformed
Schwarzschild black hole from the de Sitter gauge theory formulation 
of gravity  via    Dunkl generalized derivatives has been provided\cite{9}. In particular, the  associated black hole   solution has been found by solving  the Einstein  field equations. Such a work, dealing with the thermodynamical properties,   has been extended  to consider  other behaviors including   the   phase transition and   the light shadow aspects \cite{10,11}. Among others,  appropriate   constraints have been imposed on the   Dunkl  deformation parameter  using  empirical 
observational data of the supermassive black holes captured by Even Horizon Telescope (EHT) \cite{20,21,22}. Recently, in the case of charged Dunkl black holes, further developments have been proposed where the shadow properties have been analyzed using CUDA-based high-performance numerical  computations  explored in machine learning techniques. With the help of such accelerated computations, the horizon radius behavior and the regions of the  parameter space  supplying  physical solutions have been  investigated. Concretely, certain  algorithms have been exploited to efficiently identify patterns in the large datasets generated by GPU-accelerated simulations, enabling a more systematic exploration of the  black hole parameter space.  In this way, strict constraints on the Dunkl deformation parameters have been derived to establish a consistent link with the shadow observations provided by  EHT \cite{cuda1}.

Motivated by such  investigations, we would like to contribute to these activities by constructing a   new black hole solution on  deformed space-time geometries. Inspired by non-commutative geometry in string theory, we propose extended derivatives  by incorporating  a real antisymmetric tensor of rank 2 carrying similarities  with   stringy  fields. Using  gauging gravity scenarios, we  first   obtain the corresponding  black hole solution by solving  the  Einstein  field equations. Then, we examine    the thermodynamic properties by computing the  relevant quantities needed to  approach    the stability,   the $P$-$V$  criticality,  and   the phase transition behaviors.   With the help of CUDA numerical approaches  exploited  in machine learning computations,  we provide  strict constraints on the deformed parameter $B$ and the charge $Q$. Precisely, we show that  there are  critical  ranges for  such  parameters where the obtained black hole behaves like  the Van der Waals  system.

The outline of this investigation is as follows. Section 2 concerns   a discussion for a systematic review on   the gauge theory formulation of   gravity via the de Sitter group theory. 
In section 3, we  build deformed charged black holes from such a  formulation via new derivatives inspired by non-commutaitve geometry in string theory. In section 4, we compute  and analyze  the  thermodynamic quantities associated with such a 
solution in order to  investigate the
stability behaviors.  In  section 5, we  approach  the
 $P$-$V$   criticality and the phase transitions using CUDA numerical computation  techniques. The last section  involves 
concluding remarks.

\section{ Gauge theory  modeling of gravity}
In this section, we review    gauge theory formulations of gravitation \cite{12,13,14,15,16,17,19,34}.   A close inspection shows that several works have developed  such a subject using different scenarios.  A priori, these are many roads to achieve such  formulations. However,  a  possible  one is to consider de Sitter group SO(4,1) as a gauge symmetry on a Minkowski  space-time with the spheric geometry endowed with the following metric 
\begin{eqnarray}
ds^2=dr^2+r^2(d\theta^2+\sin^2\theta d\phi^2)-  dt^2
	\end{eqnarray}
where one has  considered $c=1$.  In this situation, the de Sitter group involves  10 infinitesimal generators denoted by   $X_A$ satisfying  the Lie algebra relations 
 \begin{eqnarray}
\left[X^A,X^B \right]=i f^{AB}_CX^C,      \qquad  A, B,C= 1,\ldots, 10,
 \end{eqnarray}
 where  $f^{AB}_C$ are called structure constants. This has been considered as a gauge symmetry of gravitation with a gauge field $ h^A_\mu(x)$ producing a strength  field associated with
  \begin{eqnarray}
F^{A}_{\mu\nu}= \partial_\mu  h^A_\nu-\partial_\nu  h^A_\mu+ f^{AB}_C   h^B_\mu  h^C_\nu.
 \end{eqnarray}
 Changing the index notation 
   \begin{eqnarray}
A\equiv a=0,1,2,3, \qquad  [ab]= [01], [02], [12], [13], [23]
 \end{eqnarray}
  one uses 
  \begin{eqnarray}
X_a=P_a, \qquad  X_{[ab]}= M_{ab},
 \end{eqnarray}
where $P_a$ and  $M_{ab}$ are the translation and the Lorenz transformation generators, respectively. In this way, the constants of  the Lie algebra  structure are given by 
 \begin{eqnarray}
\begin{aligned}
f_{b c}^a & =f_{c[d e]}^{[a b]}=f_{[b c][d e]}^a=0 ,\\
f_{c d}^{[a b]} & =4 \lambda^2\left(\delta_c^b \delta_d^a-\delta_c^a \delta_d^b\right)=-f_{d c}^{[a b]}, \\
f_{b[c d]}^a & =-f_{[c d] b}^a=\frac{1}{2}\left(\eta_{b c} \delta_d^a-\eta_{b d} \delta_c^a\right), \\
f_{[a b][c d]}^{[e f]} & =\frac{1}{4}\left(\eta_{b c} \delta_a^e \delta_d^f-\eta_{a c} \delta_b^e \delta_d^f+\eta_{a d} \delta_b^e \delta_c^f-\eta_{b d} \delta_a^e \delta_c^f\right)-e \longleftrightarrow f,
\end{aligned}
 \end{eqnarray}
 where  $\lambda$ is a deformation parameter.   The gauge fields $h^A_\mu(x)$ will be denoted  by 
  $e^a_\mu (x)$  for  $A = a$, and by $ \omega^{ab}(x)= - \omega^{ba}$  for  $A = [ab]$.   Using such identifications, the strength  tensor field 
components are split as follows
    \begin{eqnarray}
\begin{aligned}
F_{\mu \nu}^a= & \partial_\mu e_\nu^a-\partial_\nu e_\mu^a+\left(\omega_\mu^{a b} e_\nu^c-\omega_\nu^{a b} e_\mu^c\right) \eta_{b c}=T_{\mu \nu}^a, \\
F_{\mu \nu}^{a b}= & \partial_\mu \omega_\nu^{a b}-\partial_\nu \omega_\mu^{a b}+\left(\omega_\mu^{a c} \omega_\nu^{d b}-\omega_\nu^{a c} \omega_\mu^{d b}\right) \eta_{c d}  -4 \lambda^2\left(e_\mu^a e_\nu^b-e_\nu^a e_\mu^b\right)=R_{\mu \nu}^{a b}.
\end{aligned}
 \end{eqnarray}
 In this analysis,  the quantity  $F^{a}_{\mu\nu}=T^{a}_{\mu\nu}$  has been   considered  as the torsion tensor. However,  the tensor $F^{ab}_{\mu\nu}=R^{ab}_{\mu\nu}$  has been interpreted   as
the curvature tensor of a Riemann–Cartan space–time  corresponding to  the gravitational gauge fields   $e^a_\mu (x)$ and  $ \omega^{ab}(x)$.   According to  \cite{15},  the  action describing the dynamics of the gravitational gauge fields $e^{\mu}_{\nu}(x)$ and $ \omega^{ab}_{\mu} $  reads as 
\begin{equation}
S_g=\frac{1}{16 \pi G} \int \mathrm{~d}^4 x e \tilde{R}, \label{1}
\end{equation}
where  one has used \( e = \det(e_{\mu}^{a}) \) and $
\tilde{R} = \tilde{F}_{\mu\nu}^{ab} \bar{e}_{a}^{\mu} \bar{e}_{b}^{\nu}.$    In this  way, \(\bar{e}_{a}^{\mu}(x)\) denotes  the inverse of \(e_{\mu}^{a}(x)\)  satisfying  the ordinary relations
\begin{equation}
e_{\mu}^{a} \bar{e}_{b}^{\mu} = \delta_{b}^{a}, \quad e_{\mu}^{a} \bar{e}_{a}^{\nu} = \delta_{\mu}^{\nu}.
\end{equation} 
Assuming that the source of gravitation also generates an electromagnetic field  $A_{\mu}(x)$,  the  corresponding action integral  takes the following form
 \begin{equation}
  S_{em} = -\frac{1}{4Kg^2} \int d^4x \, e \, A_\mu^a \, \overline{A}_a^\mu, \label{2}
\end{equation} 
where \(A_\mu^a\)  and \(\overline{A}_a^\mu\) are  given by 
\begin{equation}
A_\mu^a = A_\mu^v e_v^a, \quad A_\mu^v = \overline{e_a}^v e_b^b \eta^{ab} A_{\mu\rho}, \quad \overline{A}_a^\mu = A_\mu^v \overline{e}_a^v.
\end{equation}
  $K$ is a constant that  could be  taken  in a convenient form  in order   to simplify  the solutions of the field equations and
 $A_{\mu \rho} = \partial_\mu A_\rho - \partial_\rho A_\mu.$  \( g \) denotes  the gauge coupling constant \cite{12,13,14,15,16,17,19,34}.   Adding up   Eq.(\ref{1}) and  Eq.(\ref{2}),  one obtains the action  total integrals  associated with  the system composed of the two fields which  can be expressed as 
\begin{equation}
S = \int d^4 x \left[ \frac{1}{16 \pi G} \tilde{R} - \frac{1}{4 K g^2} A_\mu^a \overline{A}_a^\mu \right] e.
\end{equation}  
Using the variational principle \cite{12,13,14,15,16,17,19,34}, the  field equations for the gravitational potentials \( e^a_\mu(x) \)   are given by 
\begin{equation}
\tilde{R}^a_\mu - \frac{1}{2} \tilde{R} e^a_\mu = 8\pi G \tilde{T}^a_\mu, 
\end{equation}
were  one has used 
\begin{equation}
\tilde{R}^a_\mu = \tilde{R}^{ab}_{\mu\nu} \overline{e}^\nu_b.
\end{equation}
 \(\tilde{T}^a_\mu\) is the energy--momentum tensor of the electromagnetic field  which reads as 
\begin{equation}
\tilde{T}^a_\mu = \frac{1}{K g^2} \left( A^b_\mu A^a_\nu \overline{e}^\nu_b - \frac{1}{4} A^b_\nu A^b_b e^a_\mu \right). 
\end{equation}
The field equations for the other gravitational gauge potentials \(\omega^{ab}_\mu(x)\) are equivalent to 
$\tilde{R}^a_{\mu\nu} = 0$.

Having reviewed the  needed backgrounds, we move to build a new class of charged  black hole solutions by introducing new derivatives inspired by non-commutative geometry in string theory.
 
  \section{Deformed  charged black holes}
Borrowing ideas from   stringy non-commutative geometry  and  motivated by  Dunkl formalism computations, we  would like to elaborate  a class  of charged black holes in the context  of  deformed space-times. Precisely, we   provide  a  space-time extension  by incorporating   new derivatives in the black hole  building models from the  de Sitter  gauge theory formulation of  gravitation \cite{12,13,14,15,16,17,19,34}. Roughly,  we propose the following generalization of the standard derivatives in cartesian coordinates
\begin{eqnarray}\label{}
	D_{x_{\mu}}= \frac{\partial}{\partial x_{\mu}}+ B_{\mu\nu}x^\nu
	\end{eqnarray}
where  $B_{\mu\nu}$ is  a  real  tensor  of  rank 2 which will be  fixed later on. Before going ahead,  such derivatives could  find a place in the  study of  non-commutative geometry in string theory with a nonzero  antisymmetric tensor called B-field   by turning the star product  computations\cite{35}. In particular, this could be  linked by considering   the   mapping 
\begin{eqnarray}\label{}
 B_{\mu\nu} \to i  B_{\mu\nu}
	\end{eqnarray}
where the  involved deformed parameters   could  be identified with the B-field  degree of freedoms  of  the superstring theory spectrum in the presence of D-brane objects.  Forgetting about  such links with  the  stringy non-commutative geometry and the associated techniques including the  star  product,    we consider  a special extension of usual   derivatives 
\begin{eqnarray}\label{}
D_{t}&=&\frac{\partial}{\partial t},\\
	D_{x_{i}}&=& \frac{\partial}{\partial x_{i}}+ B_{ij}x^j \quad\quad i,j=1,2,3
	\end{eqnarray}
where now   $B_{ij}$ is  a  real  tensor  of  rank 2	 in three dimensions.  A priori, many forms of such a tensor  could be considered and examined. However, an economical one corresponds to a 
antisymmetric  tensor with the following 
 matrix representation.  
\begin{eqnarray}
 B_{ij}=
\begin{pmatrix}
0 & B_1 & B_2\\
-B_1 & 0 & B_3\\
-B_2& -B_3 & 0
\end{pmatrix}.
 \end{eqnarray}
 For simplicity reasons facilitating  the computations of the  deformed   charged black hole quantities  that we are after,   we consider  a particular case where   we have  chosen 
 \begin{eqnarray}
  B_1 = B_2= B_3=B.
 \end{eqnarray}
 In this way,  we have 
\begin{eqnarray}
 B_{ij}=B \epsilon_{ij},
 \end{eqnarray}
where $\epsilon_{ij}$ is the usual epsilon tensor being completely antisymmetric 
 \begin{eqnarray}
 \epsilon_{ij} = -\epsilon_{ji}.
 \end{eqnarray}
$B$  is  now a   real  free  parameter of dimension $ [L^{-2}] $ being relevant   in the present investigation.   At this level, one could bring  certain comments.   First,  it  is  noted in passing that the singular behavior of such a matrix poses no problem in the following calculations.  In addition, non-invertible matrices have a number of applications in physics for understanding the behavior of certain systems involving    conserved quantities.   Second,   extra constraints on the  deformed parameter  $B$  could be imposed when one   approaches the physical properties including  either thermodynamics and  or optics via the  falsification mechanism.   Exploiting the spherical coordinates,  the above derivatives  can be expressed as 
	\begin{equation}\label{am}
	\begin{aligned}
	&D_{t}=\frac{\partial}{\partial t}\\
	&D_{r}=\frac{\partial}{\partial r}\\
	&D_{\theta}=\frac{\partial}{\partial \theta}+Br^2(\cos \phi+\sin \phi),\\
	&D_{ \phi}=\frac{\partial}{\partial \phi}  +Br^2\sin\theta\left(\cos \theta(\cos \phi-\sin \phi)-\sin\theta\right).
	\end{aligned}
	\end{equation}
 Having introduced such derivatives,  we construct a family of deformed  charged black holes from de Sitter gauge theory formulation of gravitation.  We assume that the deformed spacetime metric is static, spheric and
symmetric with the following  line element 
 \begin{eqnarray}\
		ds^{2}=f_d(r) dt^{2}-\frac{1}{f_d(r)}dr^{2}-r^{2}\left(d\theta^{2}+\sin^{2} \theta d \phi^{2}\right),
		\end{eqnarray}
corresponding to  the   non-zero components   of the  involved  metric $g_{\mu \nu}$.  The  function  $f_d(r)$  will carry information on the  proposed  deformed space-time geometry.
 According to  \cite{ 36,37}, we consider  a particular form  for spherically gravity fields, created by a point-like mass $M$,  given by  the  following ansatz 
	\begin{align}\label{am}
	e_\mu^0=(A, 0,0,0), \;  \;e_\mu^1=\left(0, \frac{1}{A}, 0,0\right), \;\;
	e_\mu^2=(0,0, r\, C, 0), \;\; e_\mu^3=(0,0,0, r\, C \sin \theta),
	\end{align}
	in addition to the following spin connections 
	\begin{align}\label{am}
	\omega_\mu^{01}=(U, 0,0,0), \;  \; \omega_\mu^{12}=(0,0, W, 0), \;  \; \omega_\mu^{13}=(0,0,0, Z \sin \theta), \;  \;
	\omega_\mu^{23}=(V, 0,0, \cos \theta),
	\end{align}
	where $A, C, U, W, Z$ and $V$ are radial  functions specified later on.   Indeed, such functions  should satisfy certain field equations of motion obtained from  the Einstein equations  given by
	\begin{eqnarray}\label{eeq}
	\tilde{R}_{\mu}^{\nu}-\frac{1}{2}e_{\mu}^{\nu}\tilde{R}=8\pi G \tilde{T}^\nu_\mu,
	\end{eqnarray}
 where $\tilde{R}_{\mu}^{\nu}$ and $\tilde{R}$ denote  the Ricci tensor and the Ricci scalar, respectively.  $\tilde{T}^\nu_\mu$ represents  the energy-momentum tensor of the electric field   which reads  
\begin{eqnarray}\label{am}
	\tilde{T}^\nu_\mu= \frac{1}{K g^{2}} \left( A_{\mu}^{b} \, A_{a}^{\nu} \, \overline{e}_{b}^{a} - \frac{1}{4} A_{a}^{b} \, A_{b}^{a} \, e_{\mu}^{\nu} \right).
	\end{eqnarray}
	We need to calculate the non-null components
 of $\tilde{R}^{a}_{\mu \nu}$ and $\tilde{R}^{ab}_{\mu \nu}$.  For the first tensor, we find 
\begin{equation}
\begin{aligned}
& \tilde{R}_{01}^0=-\frac{A A^{\prime}+U}{A}, \quad \tilde{R}_{03}^2=-r C V \sin \theta, \quad \tilde{R}_{12}^2=C+r C^{\prime}-\frac{W}{A}, \\
& \tilde{R}_{02}^3=r C V, \quad \tilde{R}_{13}^3=\left(C+r C^{\prime}-\frac{Z}{A}\right) \sin \theta.
\end{aligned}
\end{equation}	
Handling the second one,  we obtain
\begin{equation}
\begin{aligned}
& \tilde{R}_{01}^{01}=-U^{\prime}-4 \lambda^2, \quad \tilde{R}_{01}^{23}=-V^{\prime}, \\ &\tilde{R}_{23}^{13}=(Z-W-Br^2 (\cos\phi +\sin\phi)\tan\theta) \cos \theta, \\
& \tilde{R}_{02}^{02}=-W U-4 \lambda^2 r C A, \quad \tilde{R}_{02}^{13}=V W, \\
& \tilde{R}_{03}^{03}=-\left(Z U+4 \lambda^2 r C A\right) \sin \theta, \quad \tilde{R}_{12}^{12}=W^{\prime}-4 \lambda^2 r \frac{C}{A}, \\
& \tilde{R}_{03}^{12}=-V Z \sin \theta, \quad \tilde{R}_{13}^{13}=\left(Z^{\prime}-4 \lambda^2 r \frac{C}{A}\right) \sin \theta, \\
& \tilde{R}_{23}^{23}=-\left(1-Z W+4 \lambda^2 r^2 C^2 -Br^2(\cos\phi +\sin\phi)\cot\theta \right) \sin \theta .
\end{aligned}
\end{equation}	
The non–null  components of  the  Ricci tensor  are found to be 
 \begin{equation}
\begin{aligned}
\tilde{R}_0^0= & -U^{\prime}-\frac{(W+Z) U}{r A C}-12 \lambda^2, \\
\tilde{R}_1^1= & -U^{\prime}+\frac{\left(W^{\prime}+Z^{\prime}\right) A}{r C}-12 \lambda^2, \\
\tilde{R}_2^1= & \left(Z-W-Br^{2}(\cos\theta + \sin\theta)\cot\theta \right) \frac{A \cos \theta}{r C \sin \theta}, \\
 \tilde{R}_2^2=&\frac{W Z-1+Br^2(\cos\phi +\sin\phi)\cot\theta}{r^{2}C^2} +
\frac{W^{\prime} A}{r C}-\frac{U W}{r C A} -12 \lambda^2, \\
\tilde{R}_3^3=&\frac{W Z-1+Br^2(\cos\phi +\sin\phi)\cot\theta}{r^{2}C^2} +
\frac{Z^{\prime} A}{r C}-\frac{U Z}{r C A} -12 \lambda^2.
\end{aligned}
\end{equation} 	
In order to establish the Einstein equations, we  calculate  the Ricci scalar. Indeed,  we get 	
\begin{equation}
\tilde{R}=-2\left( U^\prime+U\frac{Z+W}{ArC}-\frac{A}{rC}(W^\prime+Z^\prime)+\frac{1-WZ-Br^2(\cos\phi+\sin\phi)\cot\theta}{r^2C^2}+24\lambda^2\right).
\end{equation}	
Assuming  that the electromagnetic field $A_\mu$ is created   by a constant electric charge $Q$ with the form  $A_\mu= (\frac{Q}{4\pi \epsilon_0 r}, 0,0,0)$. In what follows, we adopt geometrized units, in which 
\( G = c = 1 \) and \( 4 \pi \epsilon_0 = 1 \). In this convention, all 
physical quantities are expressed in  length powers. In this way,  the electromagnetic 
potential   reduces  to 
\begin{equation}
A_\mu = \left( \frac{Q}{r},\, 0,\, 0,\, 0 \right).
\end{equation}
 The non-zero components of the energy–momentum tensor  $T^a_\mu$    for the electromagnetic field sourced  by the constant electric charge  are found to be 
\begin{equation}	
\begin{aligned}
\tilde{T}_0^0 &= \frac{1}{Kg^2}\frac{AQ^2}{32\pi^2\varepsilon_0^2 r^4}, & \tilde{T}_1^1 &= \frac{1}{Kg^2}\frac{Q^2}{32A\pi^2\varepsilon_0^2 r^4}, \\
\tilde{T}_2^2 &= -\frac{1}{Kg^2}\frac{Q^2}{32\pi^2\varepsilon_0^2 r^3}, & \tilde{T}_3^3 &= -\frac{1}{Kg^2}\frac{Q^2}{32\pi^2\varepsilon_0^2 r^3}\sin\theta,
\end{aligned}
\label{K}
\end{equation}	
Putting these equations in Eq.(\ref{eeq}) and  performing analytic computations, we get the relevant   field equations of motion
\begin{eqnarray}\label{am}
	\frac{\left(1-W Z-Br^2(\cos \phi+\sin \phi) \cot \theta)\right)}{r^2 C^2}-\frac{\left(W^{\prime}+Z^{\prime}\right) A}{r C}+12 \lambda^2-\frac{Q^2}{r^4}=0,\\
	\frac{(W+Z) U}{r C A}+\frac{1-W Z-Br^2(\cos \phi+\sin \phi) \cot \theta}{r^2 C^2}+12 \lambda^2-\frac{Q^2}{r^4}=0,\\
	U^{\prime}+\frac{Z U}{r C A}-\frac{Z^{\prime} A}{r C}+12 \lambda^2+\frac{Q^2}{r^4}=0,\\
	U^{\prime}+\frac{W U}{r C A}-\frac{W^{\prime} A}{r C}+12 \lambda^2+\frac{Q^2}{r^4}=0,\\
	(W-Z)A- A\frac{Br^2 (\cos \phi+ \sin \phi) }{r C}=0.
	\end{eqnarray}
We chose the constant $K$ in Eq. (\ref{K})  such that $\frac{G}{4\pi K g^2 \varepsilon_0^2} =1$\cite{12}.  Imposing the null-torsion constraint  $F_{\mu \nu}^{a}=0$, we  obtain  the  following relations
	\begin{align}\label{15}
	U=-A A^{\prime},\quad W=Z=A,\quad V=0,\quad C=1.
	\end{align}
	In order to approach  some physical behaviors of  the underlying black hole solutions,  we consider the ansatz  $  \cot \theta(\cos \phi+\sin \phi) =1$. This particular constraint does not completely remove the angular dependence introduced by the extended derivative operators.
On the contrary, the parameter $B$ can effectively absorb the term $ \cot \theta(\cos \phi+\sin \phi) $, enabling the metric to preserve an explicitly spherical symmetric form.
   Combining such solutions and constraints, certain field equations  are  identical   producing  two  field equations involving  only one   unknown radial  function  $A$.   These two equations  read as 
	\begin{equation}\label{am}
	\begin{aligned}
	- 2 \frac{A A^{\prime}}{r}+\frac{1-A^2-Br^2}{r^2 }+12 \lambda^2 &= \frac{Q^2}{r^4},\\
	- 2 \frac{ A A^{\prime}}{r}+U^{\prime}+12 \lambda^2 &=-\frac{Q^2}{r^4}.
	\end{aligned}
	\end{equation}
	These two equations agree when we take 
\begin{eqnarray}\label{am}
	 \frac{1-A^2-Br^2}{r^2 }-U^{\prime}=  \frac{2 Q^2}{r^4}.
	\end{eqnarray}	
 Considering $ U= -\frac{(A^2){^{\prime}}}{2 }$, we get the   field equation 
	\begin{eqnarray}\label{18}
	r^{2}\left(A^{2}\right)''-2 A^{2}- 2Br^2+2= \frac{4Q^2}{r^4}.
	\end{eqnarray}
	This differential equation   can be  solved as follows
	  \begin{equation}\label{am}
	A^2=1+\frac{c_1}{r}+c_2r^2+\dfrac{2 Br^2}{3} \log \left( \dfrac{r}{r_0}\right) + \frac{Q^2}{r^2},
	\end{equation}
 where $ c_1$ and $c_2 $ are integration  constants, and $r_0$ is a constant with a length dimension, implemented   to  recover the  dimensionless logarithmic function, which we can take  to be equal to one without loss of generality.  According  to \cite{371,370},  however, different scenarios  associated with such    a logarithm  dimensionally undefined have been  approached   including the  introduction of  radial cutoffs. 
 
 In order to establish a link with certain known solutions relating to black holes, we can consider
 \begin{equation}\label{am}
c_1= -2M, \qquad c_2= -\frac{\Lambda}{3},
	\end{equation}
 where   $\Lambda$ is the cosmological constant. Taking  $B=0$,  indeed, we obtain the  solution 
 \begin{equation}\label{am}
	A^2=1-\frac{2M}{r} -\frac{\Lambda}{3}r^2+   \frac{Q^2}{r^2},
	\end{equation} 
 representing the charged de Sitter black hole found  in  \cite{38}.    The relevant    line element function is found to be 
 	\begin{eqnarray}\label{22}
	f_d(r)= A^2= f(r)+Bg(r),
	\end{eqnarray}
where one has used 
	\begin{eqnarray}\label{22}
	f(r)=1-\frac{2M}{r} -\frac{\Lambda}{3}r^2+   \frac{Q^2}{r^2}, \qquad g(r)=\dfrac{2 r^2\log r}{3}.
	\end{eqnarray}
	In this level, we would like to add a comment. In large $r$ limit, the $B r^2 \log r$ term dominates, leading to non-flat and non-dS solutions. As far as we are concerned, this behavior could find a place within non-trivial gravity theories supported by antisymmetric tensor fields belonging to the NS–NS sector of superstring  spectrums.  This may open a  way to reconsider  the study of stringy  black holes relaying  on the introduction of such tensor fields in the underlying  four dimensional action. We believe that these ideas deserve further consideration and exploration, which could be  a possible task of future works.

\begin{figure}[h]
\centering
\includegraphics[width=0.6\linewidth]{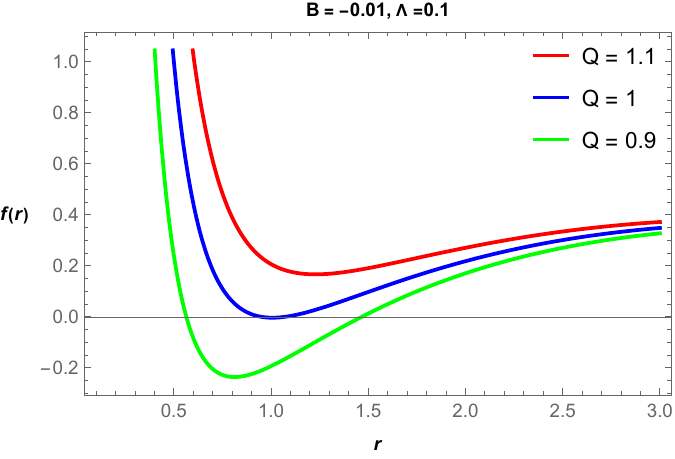}
\caption{ Metric function $f(r)$ in terms    $r$ for different values of $Q$
with $B= -0,01$ and  $\Lambda =0.1$.}
\label{F1}
\end{figure}

The obtained  black hole solution involves then  a  relevant  deformed parameter $B$, the charge $Q$ and  two integration constants $M$ and $\Lambda$, associated with the black hole  mass and the  cosmological constant, respectively.
The physical characteristics of the resulting metric will be examined in the light of a moduli space whose coordinates are the four parameters $(M,  \Lambda, Q, B)$.
In accordance with standard practices, the mass $M$  will be fixed to reduce  such  moduli space coordinates.  In this reduced moduli space, the 
charge $Q$  and the deformed parameter $B$  will be considered as  relevant  quantities whose physical significance will be analyzed up to certain requirements imposed by thermodynamical  aspects. In this   analysis, 
 the  physical significance of the corresponding solutions depends on which regions of the reduced moduli space $(B, Q)$ are taken into account.  In Fig.\ref{F1}, we present the characteristics of the obtained  solutions. 
Fixing the value of  $B $,  it has been observed that there exists a special value of the charge,
denoted by $Q_{s}$, associated with a double zero of $f(r)=0$. As is
evident from Fig.\ref{F1}, if $Q>Q_{s}$,  the solution describes a naked
singularity. For $%
Q=Q_{s}$ and $Q<Q_{s}$, however, two kinds of  solutions  could arise 
corresponding to extremal and non-extremal black holes, respectively.

\section{Thermodynamics of deformed charged black holes }

In this section, we investigate  the thermodynamics of the obtained black hole  solution  inspired by non-commutative geometry in string theory.  In particular, we are interested in the effect of the deformed parameter  $B$ on the thermal properties by computing and examining  the relevant quantities in terms of the horizon  radius $r_h$.   We start by 
 determining  the  mass  which can be  obtained from  the constraint $f(r)\bigg\rvert_{r=r_h}=0$.  Indeed,   it  is  expressed as follows
 \begin{equation}
M=
\frac{2 B \ln \! \left(r_{h} \right) r_{h}^{4}-  \Lambda r_{h}^{4} +3 Q^{2}+3 r_{h}^{2}}{6 r_{h}}
\end{equation}
where $r_{h}$ is the radius of the horizon. In the limit $B=0$, the mass reduces to 
\begin{equation}
M=
\frac{-\Lambda r_{h}^{4} +3 Q^{2}+3 r_{h}^{2}}{6 r_{h}}
\end{equation}
recovering the result obtained in \cite{39}.
The Hawking temperature
 can be computed using the relation  $T=\frac{\kappa }{2\pi }$, where $\kappa $
is the surface gravity defined by $\kappa =\frac{1}{2}\frac{\partial f(r)}{%
\partial r}\bigg\rvert_{r=r_h}$. After computations,  the 
temperature  is found to be 
\begin{equation}
T=
\frac{6 B \ln \! \left(r_{h} \right) r_{h}^{4}+\left(2 B -3 \Lambda \right) r_{h}^{4}+3 r_{h}^{2}-3 Q^{2}}{12 r_{h}^{3} \pi}.
\end{equation}
This temperature can be reduced to known expressions. 
Taking  $B=0$, for  instance, we can
recover the Hawking temperature of the  charged black holes with dS geometries
\begin{equation}
T=
\frac{ -3 \Lambda  r_{h}^{4}+3 r_{h}^{2}-3 Q^{2}}{12 r_{h}^{3} \pi}.
\end{equation}
For  $\Lambda=Q= 0$, moreover,   we get  the well-known temperature of the Schwarzschild black
hole given by  $T=\frac{1}{4\pi r_{h}}$ \cite{100}.
\begin{figure}[h!]
\centering
\includegraphics[scale=0.8]{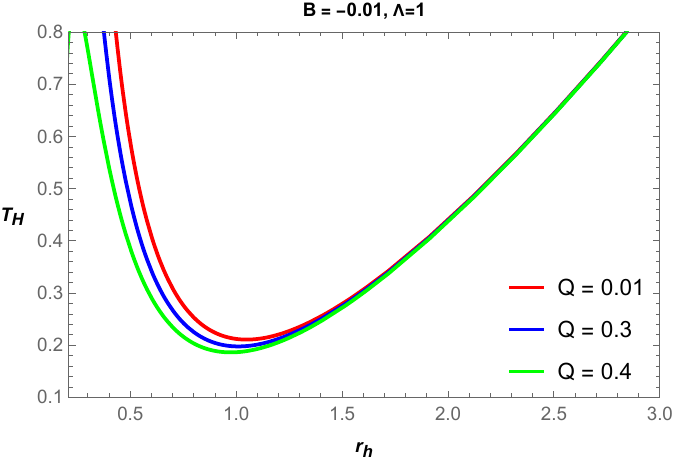}
\caption{Hawking temperature in terms of $r_{h}$  by  varying  $Q$
for $B=-0,01$ and $ \Lambda=1$.}
\label{F2}
\end{figure}
 In order to illustrate the behavior of the temperature, We can plot the
above expression as a function of the radius of the event horizon, limiting ourselves to the appropriate regions
of reduced modular space. Fig.\ref{F2} depicts such
 thermal behaviors.
It has been observed  that the Hawking temperature decreases  to a minimal  value.
After that, it increases. A close inspection reveals that
the minimal value increases when the charge $Q$ decreases.

Exploiting  the first-law of
the thermodynamics,  we can compute the entropy   via the relation $dm=TdS$  leading to \begin{equation}
S=\int_0^{r_h} \frac{1}{T} \frac{\partial m}{\partial r} dr.
\end{equation}
Integrating such a relation, we obtain 
\begin{equation}
S=\pi r^2_{h}.
\end{equation}

Before going to the main objective,  we  would like to   inspect  the stability behaviors.  Concretely, we discuss  both global and local stabilities  by evaluating the relevant quantities within the thermodynamic formalism. First, we  examine the  global stability by considering the Gibbs free energy, defined as follows 
\begin{equation}
G = M - T_H S,
\end{equation}
which leads to the following expression
\begin{equation}
G = \frac{-2 B \ln(r_h) r_h^4 + (-2 B + \Lambda) r_h^4 + 3 r_h^2 + 9 Q^2}{12 r_h}.\label{G}
\end{equation}
For $B = 0$,  it has been observed that the Gibbs free energy reduces  to
\begin{equation}
G = \frac{\Lambda r_h^4 + 3 r_h^2 + 9 Q^2}{12 r_h},
\end{equation}
recovering the result found in \cite{39}.   It is denoted that  a thermodynamic system is globally stable when the Gibbs free energy is negative $(G<0)$. Conversely, a positive Gibbs free energy $(G>0)$ indicates a thermodynamically unstable state.
\begin{figure}[!h]
\centering
\begin{tabular}{cc}
\includegraphics[width=7.5cm, height=7cm]{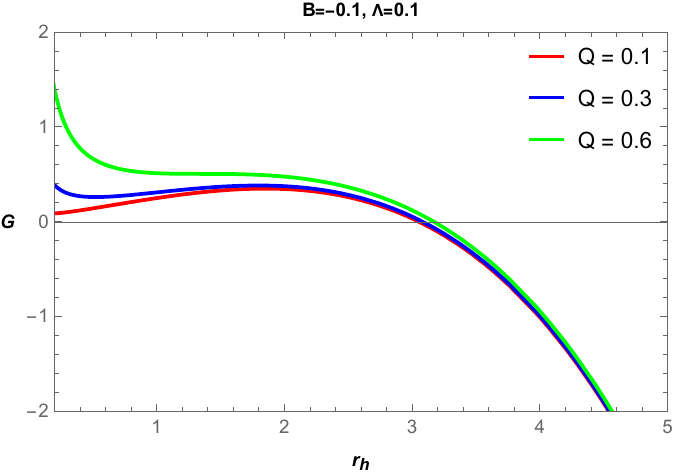} &
\includegraphics[width=7.5cm, height=7cm]{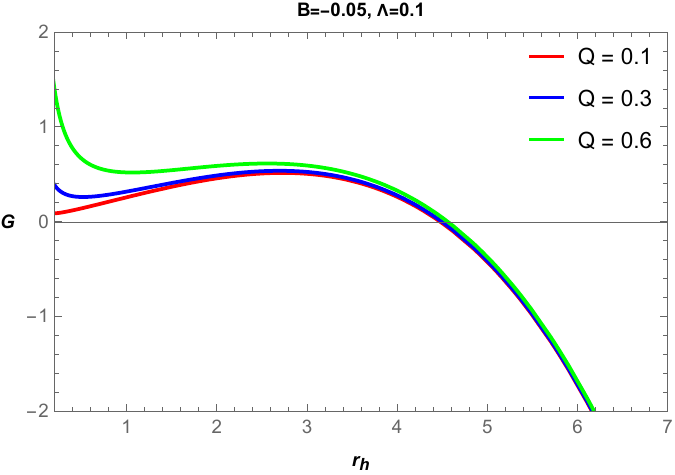}
\end{tabular}
\caption{Gibbs free energy as a function of $r_h$ with $\Lambda=0.1$.}
\label{F33}
\end{figure}
To examine these behaviors, Fig.~\ref{F33} shows the variation of $G$ as a function of $r_h$ within certain acceptable regions of the moduli space.  
For large horizon radius values,  the Gibbs free energy vanishes at a specific value $r_h = r_h^0$.  
When $r_h > r_h^0$, $G$ becomes negative, indicating that the system is globally stable. Otherwise, the system is globally unstable.

\begin{figure}[!h]
\centering
\begin{tabular}{cc}
\includegraphics[width=7.5cm, height=7cm]{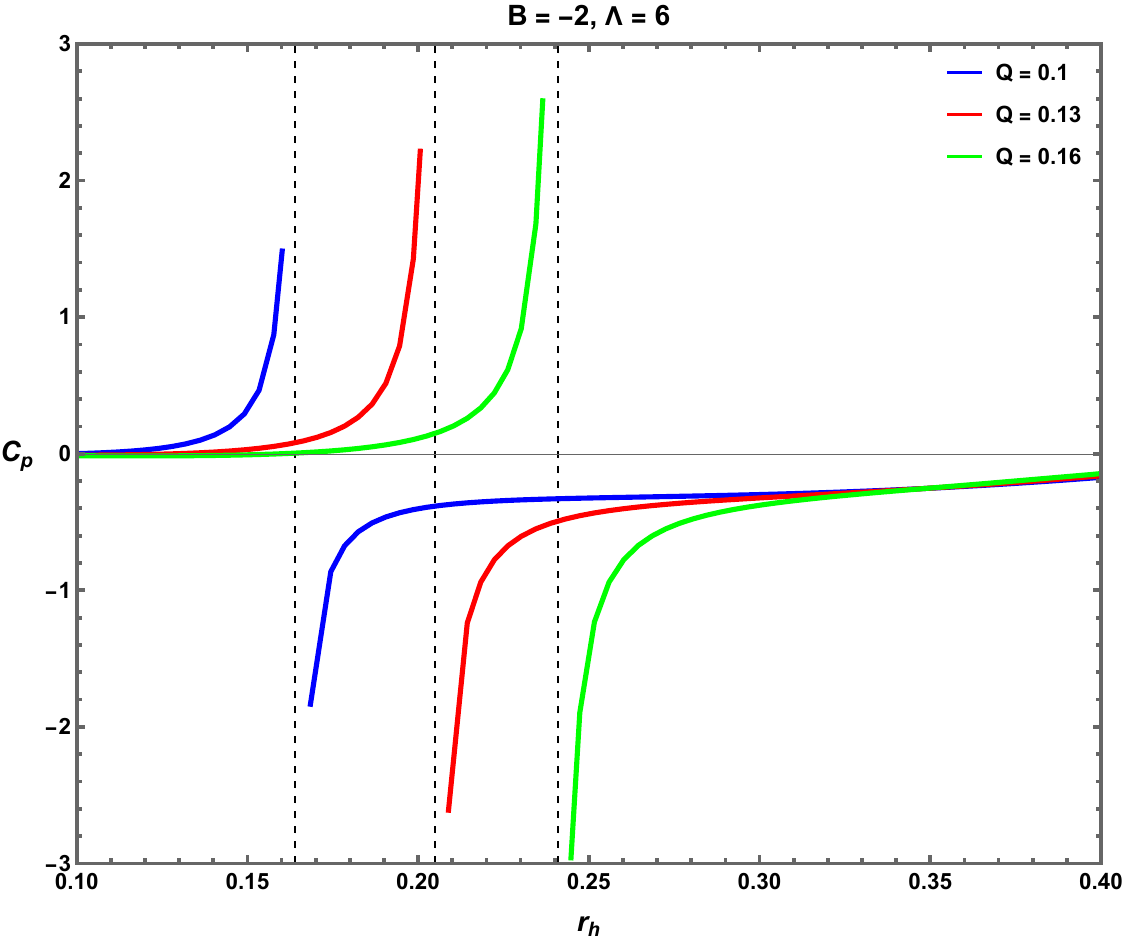} &
\includegraphics[width=7.5cm, height=7cm]{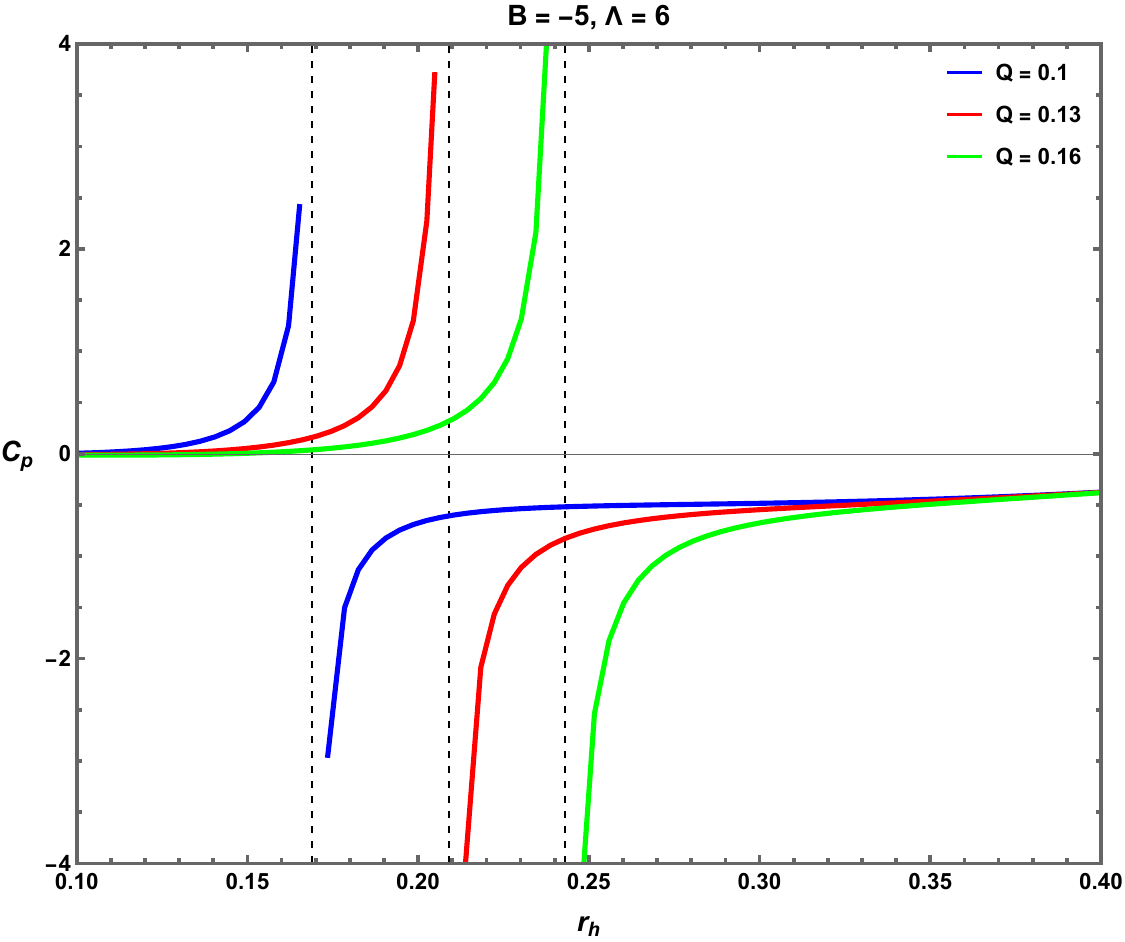}
\end{tabular}
\caption{ Heat capacity in  terms  of $r_{h}$ for different values of $Q$.}
\label{F3}
\end{figure}

In order to approach the local stability behaviors, one needs to determine the heat capacity via the relation  $C_{p}=(\frac{\partial m}{\partial r_h})/(\frac{\partial T}{\partial
r_{h}})$.  The computations provide 
\begin{equation}
C_p=\frac{2 \pi  r_{h}^{2} \left(6 B  r_{h}^{4}  \ln \! \left(r_{h} \right)+\left(2 B -3 \Lambda \right) r_{h}^{4}+3 r_{h}^{2}-3 Q^{2}\right) }{6 B  r_{h}^{4} \ln \! \left(r_{h} \right) +\left(8 B -3 \Lambda \right) r_{h}^{4}-3 r_{h}^{2}+9 Q^{2}}.
\end{equation}
For $Q=B=0$,    we  recover  the standard dS-Schwarzschild black hole expression. It has been shown that the sign of the  heat capacity can be used to control stability 
of the corresponding black hole solutions. Indeed,  a locally stable thermodynamical
system corresponds to $C_{p}>0$, whereas an unstable solution can occur  if $C_{p}<0$. 
In Fig.\ref{F3},  we plot $C_{p}$ as
a function of $r_{h}$.

It has been remarked that  the heat capacity curves  exhibit discontinuous  behaviors at  certain critical   horizon radius
$r_{h}=r_{h}^{c}$ where the temperature takes  critical values   at the used points in the reduced
moduli space.  In such points, the heat capacity  shows  divergence behaviors.  For  $B=-5$, $\Lambda
=6$,  and $Q=0.1$, for instance,   we obtain   that $r_{h}^{c}=0.17$. Fixing  the deformed  parameter $B$, we remark  that $r_{h}^{c}$
increases by increasing the charge $Q$.  
However, $r_{h}^{c}$ decreases   by increasing the deformed parameter $B$.  Moreover,    two separated branches appear 
which show that  the obtained  black holes exhibit  stable  and  unstable  configurations corresponding to  $r<r_{h}^{c}$ and $r>r_{h}^{c}$, respectively. 
The detection of  the divergence  behaviors clearly indicates a second-order phase transition. We expect similar behaviors to be  developed in other regions of the
 reduced moduli space.

\section{$P$-$V$ criticality  and  CUDA  numerical computations }
In this section, we  study   the   critical behaviors and the phase transitions  of the obtained  deformed charged black holes in four dimensions by
 exploiting the  techniques of CUDA  computations used in machine learning  methods. Such numerical  approaches   impose  extra constraints  on the deformed parameter  $B$ recovering certain universal behaviors in the black hole thermodynamics.
 \subsection{$P$-$V$ criticality} 
\begin{figure}
\centering
\includegraphics[scale=0.85]{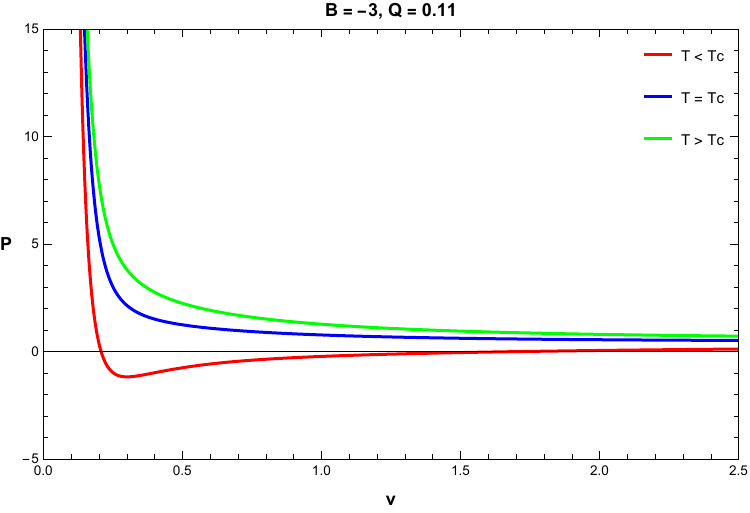}
\caption{Pressure in terms of $v$ for different values of $T$  for  $
B=-3 $ and  $Q =0,11$.}
\label{F41}
\end{figure}

\begin{figure}[!h]
\centering
\begin{tabular}{cc}
\includegraphics[width=8cm, height=8cm]{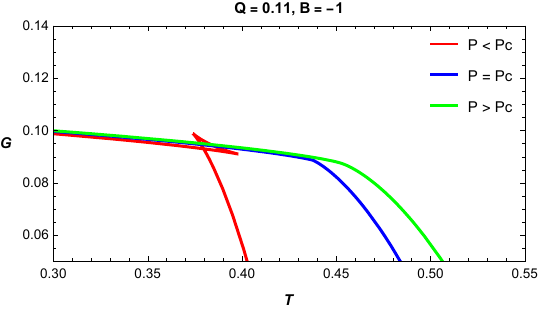} &
\includegraphics[width=8cm, height=8cm]{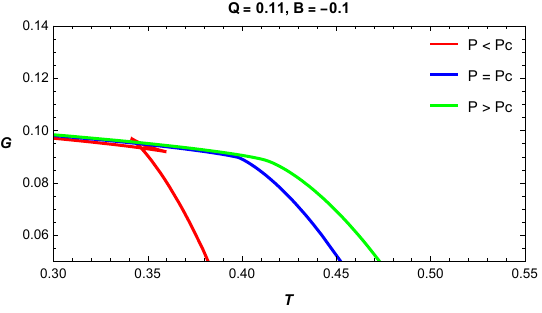}
\end{tabular}
\caption{Gibbs free energy  in terms of the  temperature for different values of  $P$, 
$B$,  and $Q$.}
\label{F5}
\end{figure}

First,  we  discuss  the $P$-$V$  criticality behaviors of the deformed charged  like-de Sitter black hole solutions.  To do so,  we establish the thermodynamic state equation  by  interpreting 
the cosmological constant $\Lambda $   as a pressure\begin{equation}
P=-\frac{\Lambda }{8\pi }, \label{PP}
\end{equation}%
where  the   corresponding conjugate variable  can be  identified with the thermodynamic
volume. Indeed, the pressure is found to be 
\begin{equation}
P=-\frac{6 B \ln \! \left(r_{h} \right) r_{h}^{4}+2 B \,r_{h}^{4}-12 \textit{T}  r_{h}^{3} \pi -3 Q^{2}+3 r_{h}^{2}}{24 r_{h}^{4} \pi}.
\end{equation}
Taking  $B=0$, $\Lambda=0$,  and $Q=0$, we get  the usual expression given by 
\begin{equation}
P=\frac{ 4\pi \textit{T}  r_{h}-1}{8 \pi r_{h}^{2} \pi}.
\end{equation}
Using the specific volume  $v = 2r_h$,   the pressure can be rewritten as follows 
\begin{equation}
P=\dfrac{B (log(2)-log(v))}{4\pi}-\dfrac{B}{12 \pi}+\dfrac{T}{v}+\dfrac{2Q^{2}}{v^{4}\pi}-\dfrac{1}{2 v^{2} \pi}.
\end{equation}
The critical points can be obtained by solving  the following constraints 
\begin{equation}
\frac{\partial P}{\partial \upsilon }=0,\hspace{1.5cm}\frac{\partial ^{2}P}{%
\partial \upsilon ^{2}}=0.
\end{equation}
Handling  such constraints,  one can calculate the the critical pressure, the  critical temperature, and  the critical volume. The computations  provide 
\begin{eqnarray}
P_c&=&\frac{3  B\left(12 B \,Q^{2}+\xi \right)  \left(\ln \! \left(2\right)-2 \ln \! \left(-\frac{\sqrt{-B \xi }}{B}\right)\right)-114 B^{2} Q^{2}-11 B \xi }{12 \xi ^{2} \pi} \nonumber\\ 
T_{c}&=&-\frac{  \sqrt{2} B\left(8 B \,Q^{2}+\xi \right) }{\xi  \sqrt{-B \xi }\, \pi}\\
v_{c}&=&-\frac{\sqrt{-2 B \xi }}{B}  \nonumber
\end{eqnarray}	
where  one has used $
\xi=\xi(B,Q)=1+\sqrt{24 B \,Q^{2}+1}$.  These computations bring  requirements on the involved parameters which will  be considered  in the discussion of the   $P$-$V$ criticality    universal  behaviors.  In this respect, the critical triple $(P_{c},  T_{c}, v_{c})$ provides the following ratio
\begin{equation}
\chi=\dfrac{P_{c}v_{c}}{T_{c}}=\frac{
(36 B Q^2 + 3\xi)\left(-\ln(2) + 2\ln\left(-\frac{\sqrt{-B \xi}}{B}\right)\right) + 114 B Q^2 + 11\xi 
}{
96 B Q^2 + 12\xi
}
\end{equation}
which  does not retain a constant value as  charged  AdS black holes \cite{40}. 
In Fig.\ref{F41},  the  $P$-$v$ criticality  diagram is illustrated.
 For  $T < T_c$,  it follows  from this figure     that  the obtained black hole system   behaves  like an extended Van der
Waals gas  with  inflection points. To investigate the associated  phase transitions, we express the  Gibbs free energy as a function of pressure. By substituting Eq.(\ref{PP}) into Eq.(\ref{G}), we obtain
\begin{equation}
G=\frac{-2 B r_h^{4} \ln \! \left(r_h \right) -\left(8 \pi P  +2 B \right) r_h^{4}+3 r_h^{2}+9 Q^{2}}{12 r_h}.
\end{equation}
Fig.\ref{F5}  illustrates  $G$  in terms  of $T$
for different  points   of the moduli space. 

From this figure, it has been   shown  the effect of the black hole parameters on
the phase transitions. Moreover, it has been remarked that  the $G-T$ curves behave similarly for
different critical values  depending on  the acceptable regions of the reduced moduli space.  For $P<P_c$, we  observe  the characteristic
swallow-tail   ensuring the existence of a phase  transition.  Similar behaviors appear  in  the Van der Waals fluid systems.
\subsection{Constraining  parameters  using CUDA  numerical computations }
\begin{figure}
\centering
\includegraphics[scale=1.2]{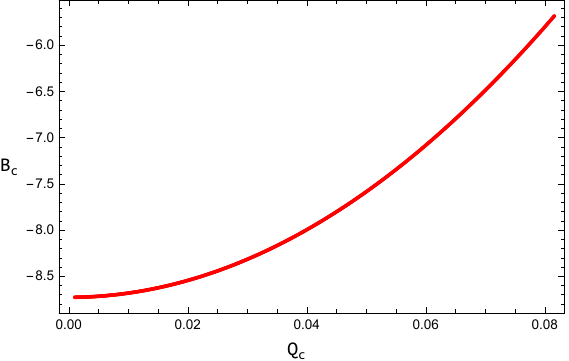}
\caption{ $B_c$   as a  function  of  $Q_c$ producing Van der Waals behaviors. }
\label{F6}
\end{figure}

In this part, we  would like to discuss  Van der Waals behavior limits of  the obtained deformed  black hole.   In   $P$-$V$ criticality studies, this   could   be ensured  by  the constraint  
\begin{equation}
\chi=\frac{3}{8}.
\end{equation}
This identification  may impose strict constraints  on the involved  parameter ranges   which will be referred to as  critical  ones. To determine the corresponding  points ($Q_c,  B_c)$  of the reduced moduli space,  we should handle the   following constraints
\begin{equation}
 B_c< 0,  \qquad    24 B_c \,Q_c^{2}+1>0, \qquad  \dfrac{P_{c}v_{c}}{T_{c}}=\frac{3}{8},
\end{equation}
where the first ones  are  compelled  by thermodynamics computations evincing non-physical quantities. An inspect   reveals  that   approaching  these  constraints may need intensive numerical computations.   However,  we  numerically investigate  such critical quantities using parallel programming techniques such as warps, shared memory and thread synchronization \cite{101}.  Precisely, we  would like to elaborate  a CUDA code capitalizing  on the GPU architecture.  It is denoted that CUDA is a general purpose parallel computing platform and programming model that leverages the parallel compute engine in NVIDIA GPUs. Precisely, the GPU architecture enables sharing the workloads between a number of streaming multiprocessors (SMs) in a highly parallel manner. In addition, these GPUs introduce a multitude of powerful methods and efficient tools to further exploit the architecture. For each  new version,  there are more interesting ways to build CUDA kernels for more powerful performance and efficiency.  Moreover, CUDA has proven to be a powerful tool in black hole research, enabling efficient GPU-accelerated simulations \cite{cuda2,cuda3,cuda4}. It reduces computation time. This not only  improves numerical stability but also allows the  theoretical predictions  including black hole physics \cite{41,42}. These techniques have been    implemented in the study of  black hole shadows needed to provide  bridges  with observational data  explored by EHT \cite{cuda1}.  Combining CUDA-accelerated simulations with machine learning techniques, large datasets generated from complex simulations can be used to train predictive models, enabling faster and more accurate analysis of black hole properties and their observational signatures. Roughly,     the  CUDA numerical method has been exploited to approach universal black hole behaviors. More precisely, we elaborate an appropriate   CUDA   numerical code allowing to  plot  $B_c$  as function of the   critical charge $Q_c$   in  Fig.\ref{F6}.
This figure provides   the allowed range of the critical  values of   $B_c$ for suitable charges  $Q_c$. Each point on   the  ($B_c-Q_c$)  elaborated curve  indicates  a specific location  $\left( Q_c,  B_c \right)$ in the  reduced moduli producing the universal  value    $ \chi=\frac{3}{8}$. At such  curve  points,  the deformed black hole behaves like  the  Van der Waals  thermodynamical system by imposing strict constraints on the associated critical  ranges.  This scenario can  be considered as an alternative approach to  constraint   black hole parameters using optical properties including the shadow and the deflection angle  computations.  This has be performed by exploiting  the observational empirical  data of M87$^*$ and SgrA$^*$ furnished by  the EHT international collaboration.
\section{Conclusions}
In this work, we have presented a   new class of black hole solutions on  deformed space-time geometries.  Inspired by non-commutative geometry in string theory, we have   considered   extended derivatives  by incorporating  a real antisymmetric tensor of rank  2 carrying similarities  with the   stringy  B-field. Using  gauging gravity scenarios, we  have constructed the corresponding  black hole solution by solving  the  Einstein  field equations. Then, we have  examined   the thermodynamical properties by computing the relevant quantities. Concretely, we have  investigated    the stability analysis and the criticality universal behaviors.   With the  help of CUDA  numerical computations  exploited  in machine learning computations,  we  have observed   similarities with  Van der Waals fluids by imposing  strict constraints on the deformed parameter $B$  and the charge $Q$. In particular, we have found  suitable ranges   where the obtained black hole behaves  like  Van der Waals fluids.  This  approach  has been interpreted  as an alternative  road to  constrain  black hole parameters using  shadow  computations by utilizing   the observational empirical  data of M87$^*$ and Sgr A$^*$  provided   by the EHT international collaboration.

This work leaves certain equations. Other behaviors could be addressed in future works including the introduction of  extra parameters associated either with internal or external  sides  of the black hole moduli space. Furthermore, another extension would be to explore
the characteristics of other properties, such as shadows, where a  potential contact with the M87$^*$ and Sgr A$^*$ bands could be
established using in-depth numerical calculations including  CUDA combined with machine learning methods.  We hope to be able to report on these open questions elsewhere.

  {\bf Data availability}\\
  Data sharing is not applicable to this article.

\section*{Acknowledgements}
 MJ would like to thank S. E. Baddis  and H. Belmahi for discussions and  scientific help on  code numerical computations.   The authors would like to thank the anonymous reviewers for their helpful comments, remarks,  and suggestions, which significantly improved the overall manuscript. This work has been  done with the support of the CNRST in the frame of the PhD Associate Scholarship Program PASS.

\end{document}